\documentclass[aps,prb,twocolumn,floatfix]{revtex4-1}

\usepackage{graphicx}
\usepackage{bm}

\def\bk{\bm k}
\def\br{\bm r}

\begin{document}
\title{Comparative study of tight-binding and \textit{ab initio} electronic structure calculations focused on magnetic anisotropy in ordered CoPt alloy}
\author{J.~Zemen,$^{1,2}$ J.~Ma\v{s}ek,$^{3}$ J.~Ku\v{c}era,$^{2}$ J.~A.~Mol,$^{4}$ P.~Motloch,$^{2}$ and T.~Jungwirth$^{2,1}$}
\affiliation{$^{1}$School of Physics and Astronomy, University of Nottingham, Nottingham NG7 2RD, UK}
\affiliation{$^{2}$Institute of Physics ASCR, v. v. i., Cukrovarnick\'a 10, 162 00 Praha 6, Czech Republic}
\affiliation{$^{3}$Institute of Physics ASCR, v. v. i., Na Slovance 2, 182 21 Praha 8, Czech Republic}
\affiliation{$^{4}$Kavli Institute of Nanoscience, Delft University of Technology, Lorentzweg 1, 2628 CJ Delft, The Netherlands}
\date{\today}

\begin{abstract}
An empirical multiorbital ($spd$) tight binding (TB) model including magnetism and spin-orbit coupling is applied to calculations of magnetic anisotropy energy (MAE) in CoPt $L1_0$ structure. A realistic Slater-Koster parametrisation for single-element transition metals is adapted for the ordered binary alloy. Spin magnetic moment and density of states are calculated using a full-potential linearized augmented plane-wave (LAPW) \textit{ab initio} method and our TB code with different variants of the interatomic parameters. Detailed mutual comparison of this data allows for determination of a subset of the compound TB parameters tuning of which improves the agreement of the TB and LAPW results. MAE calculated as a function of band filling using the refined parameters is in broad agreement with \textit{ab initio} data for all valence states and in quantitative agreement with \textit{ab initio} and experimental data for the natural band filling. Our work provides a practical basis for further studies of relativistic magnetotransport anisotropies by means of local Green's function formalism which is directly compatible with our TB approach.
\end{abstract}

\maketitle

\section{Introduction}
\label{intro}

Ordered CoPt alloys have been studied widely as they hold potential for applications in high density magnetic recording due to the combination of exchange and spin-orbit interactions giving rise to large magnetic anisotropies. Tunnelling magnetoresistance,\cite{12n_kim:2008_a} tunnelling anisotropic magnetoresistance,\cite{12n_park:2008_a} or spin pumping\cite{12n_skinner:2012_a} have been demonstrated in CoPt based devices. 
Our general objective is to develop an efficient numerical model allowing us to study ground state and in future studies also transport properties of spintronic devices based on CoPt or other intermetallic compound with large magnetic anisotropy energy (MAE). 

MAE has been calculated in CoPt $L1_0$ structure using \textit{ab inito} methods.\cite{12n_shick:2003_a,12n_shick:2006_a} The stability of various bulk CoPt structures\cite{12n_karoui:2012_a} has been studied recently also by \textit{ab inito} methods. On the other hand, quantum transport in micro-devices is typically described within the Green's function formalism assuming an expansion of the electronic states on a local basis set which facilitates partitioning of the system. The tight binding (TB) description of electronic structure provides a good foundation for subsequent simulations of magnetoresistance in tunnelling or ohmic regime as it assumes a local basis set.\cite{12n_mathon:1997_a,12n_krstic:2002_a}

TB schemes applied in modelling of magnetotrasport phenomena range from empirical or semi-empirical (charge self-consistent) models\cite{12n_mathon:2001_a,12n_autes:2008_a,12n_autes:2010_a} to tight binding linearized muffin-tin orbital (LMTO) model\cite{12n_kudrnovsky:2000_a,12n_xia:2002_a,12n_carva:2007_a} combining density functional theory (DFT) with the TB approach. 
The state-of-the-art magnetic DFT-based TB schemes\cite{12n_mceniry:2011_a,12n_urban:2011_a,12n_paxton:2010_a, 12n_frauenheim:2000_a,12n_hourahine:2010_a,12n_horsfield:2011_a} allow for simulations of complex systems (impurities, structural defects, surfaces) that are beyond practical capability of \textit{ab initio} calculations. In this work, we employ an empirical two-center Slater-Koster TB model\cite{12n_slater:1954_a} following the Harrison approach\cite{12n_harrison:1980_a} recently further developed by Shi and Papaconstantopoulos\cite{12n_shi:2004_a} to investigate MAE of a bulk ordered transition metal alloy and compare the results to density functional theory (DFT) calculations performed using the full-potential linearized augmented plane-wave (LAPW) program package WIEN2k. Thereby we test the transferability of the TB parameters\cite{12n_shi:2004_a} obtained by fitting to \textit{ab initio} band structures of single-element solids of the two atoms forming our ordered alloy. Rather than finding a new full set of TB parameters by fitting to the DFT band structure of the compound, we determine a minimal subset of interatomic parameters which influence the spin magnetic moments and projected density of states (DOS) in a transparent way and tune these parameters to improve the agreement of TB and DFT results. MAE calculated in a narrow range of refined values of the interatomic parameters is in quantitative agreement with DFT. We believe that our model and parameters provide good basis for future simulations of magnetotransport in CoPt and other multilayer structures. 


Our paper is organized as follows: The TB model including the exchange and spin-orbit interaction is introduced in Sec.~\ref{se_model}; TB parameters according to Shi and Papaconstantopoulos with our modifications are discussed in Sec.~\ref{se_parameters}; First comparison of experimental MAE,\cite{yermakov:1990_a} \textit{ab initio}, and TB results is presented in Sec.~\ref{se_comparison}; Sec.~\ref{se_refinement} describes the refinement of TB parameters towards agreement of spin magnetic moment and DOS with \textit{ab initio} results, it also presents corresponding MAE in comparison with DFT; Our work is briefly summarised in Sec.~\ref{se_summary}.

\section{Tight Binding Model}
\label{se_model}
Our TB model is a variant of the Linear Combination of Atomic Orbitals method for a periodic crystal where the basis has the form of Bloch sums of atomic like wavefunctions:
\begin{equation}\label{blochsum}
\langle \br|a\alpha \bk \rangle = \frac{1}{\sqrt{N}} \sum_{n=0}^{N-1}e^{i\bk\cdot({\bm R}_n+{\bm p}_a)}\phi_{a\alpha}(\br-{\bm R_n}-{\bm p}_a),
\end{equation}
where ${\bm k}$ is the Bloch wave vector, $a$ is the atom index in the unit cell, $\alpha$ is the atomic orbital quantum number, $N$ is the number of unit cells (or atoms if there is only one atom per unit cell), $n$ is the unit cell index, ${\bm R}_n$ is the unit cell vector, and ${\bm p}_a$ is the position vector of the atom $a$ in the unit cell. We consider an orthonormal set of wavefunctions $\phi_{a\alpha}({\bm r})$ constructed from the atomic orbitals (with the angular part expressed in terms of cubic harmonics given in Eq.~(\ref{app_charm})) using L\"owdin's orthonormalization procedure so the overlap matrix $\langle b\beta\bk|a\alpha\bk\rangle = \delta_{b\beta,a\alpha}$. Our single-electron Hamiltonian has three components as follows:
\begin{equation}\label{Ham_comp}
H = H_{band}+H_{Stoner}+H_{SO}.
\end{equation}
The first component $H_{band}$ contains the kinetic energy and a superposition of atomic potentials centered at each site:
\begin{equation}\label{potv}
V(\br) = \sum_{na}V^{at}_a(\br-{\bm R_n}-{\bm p}_a),
\end{equation}
where the indices $n$ and $a$ run through all unit cells and all atoms in a unit cell, respectively. The potentials $V^{at}_a(\br)$ centred at each site are spherically symmetric so the wavefunctions $\phi_{a\alpha}$ can be specified by the usual angular momentum quantum numbers. 

The matrix elements of the non-magnetic Hamiltonian term $H_{band}$ can be written in terms of on-site energies $\varepsilon_{a\alpha}$ and hopping integrals $E_{b\beta,a\alpha}({\bm \rho_n})$ which depend, in the two-center approximation proposed by Slater and Koster, only on the intersite position vector ${\bm \rho_n}={\bm R_n}+{\bm p}_a-{\bm p}_{b}$ as follows:
\begin{eqnarray} \label{ham_m_band}
\langle b\beta\bk|H_{band}|a\alpha\bk\rangle & = & \delta_{b\beta,a\alpha}\varepsilon_{a\alpha} + \sum_{n}e^{i\bk\cdot{\bm \rho_n}}E_{b\beta,a\alpha}({\bm \rho_n}), \nonumber \\
E_{b\beta,a\alpha}({\bm \rho_n}) & \equiv &
\int d\br\phi^{\ast}_{b\beta}(\br) H_n(\br) \phi_{a\alpha}(\br-{\bm \rho}_n), \;
\end{eqnarray}
where the Hamiltonian contribution $H_n(\br)$ accounts for interactions between atomic sites $a$ and $b$ in the same (${\bm R_n}=0$) or neighbouring unit cells (${\bm R_n} \ne 0$).

Following the TB scheme, we parametrize the Hamiltonian matrix instead of performing the integration of Eq.~(\ref{ham_m_band}). The hopping integrals $E_{b\beta,a\alpha}({\bm \rho_n})$ can be expressed in terms of Slater-Koster parameters ($V_{ss\sigma}$, $V_{sp\sigma}$, $V_{sd\sigma}$, $V_{pp\sigma}$, $V_{pp\pi}$, $V_{pd\sigma}$, $V_{pd\pi}$, $V_{dd\sigma}$, $V_{dd\pi}$, $V_{dd\delta}$, etc.) obtained usually by fitting to \textit{ab initio} calculations. See Eqs.~(\ref{app_E_Vsp}) and~(\ref{app_E_Vd}) in the Appendix for the so called Slater-Koster tables listing the $s$, $p$, and $d$ hopping integrals used in this work.

The on-site energies $\varepsilon_{a\alpha}$ could be approximated by the atomic values as in the original Harrison approach. Instead, we follow Shi and Papaconstantopoulos\cite{12n_shi:2004_a} who keep the on-site $k$-independent matrix diagonal but use $\varepsilon_{a\alpha}$ obtained by fitting to \textit{ab initio} data. Hence, both the on-site energies and Slater-Koster parameters are inputs of the model as we discuss in more detail in Sec.~\ref{se_parameters}.

Since the Hamiltonian has the same periodicity as the basis functions in Eq.~(\ref{blochsum}) it is diagonal in the $k$-vector. The dimension of our Hamiltonian matrix is given by the number of atoms in the unit cell and valence orbitals considered for each atom. We use $s$, $p$, and $d$~orbitals to model CoPt and there are only two atoms in the unit cell of the $L1_0$ structure. The sum in Eq.~(\ref{ham_m_band}) runs over a limited set of neighbouring sites due to localization of functions $\phi_{a\alpha}$ around site $a$.

The Hamiltonian described so far is non-magnetic. Now we double our Hilbert space by including the spin degree of freedom 
$|a\alpha\bk\rangle \rightarrow |a\alpha\xi\bk\rangle$ and add a $\bk$-dependent term $H_{Stoner}$ to account for the ferromagnetism in our system:
\begin{eqnarray}\label{Hfm}
  \langle b\beta\zeta\bk|H_{Stoner}|a\alpha\xi\bk \rangle & = & \frac{1}{2}\delta_{b\beta,a\alpha}I_{a\alpha}{\hat{\bm m}}\cdot{\bm \sigma}_{\zeta\xi}, \\
  & + & \frac{1}{2}\sum_{n}e^{i\bk\cdot{\bm \rho_n}}I_{b\beta,a\alpha}({\bm \rho_n}){\hat{\bm m}}\cdot{\bm \sigma}_{\zeta\xi}, \nonumber
\end{eqnarray}
where $I_{a\alpha}$ and $I_{b\beta,a\alpha}$ are the on-site and hopping Stoner parameters, respectively, ${\hat{\bm m}}$ is the magnetization unit vector constant at all sites, and ${\bm \sigma}$ is the vector of Pauli matrices. We derive our Stoner parameters from the exchange-split on-site energies and hopping integrals: $I_{a\alpha}=\varepsilon_{a\alpha\uparrow}- \varepsilon_{a\alpha\downarrow}$, $I_{b\beta,a\alpha}=E_{b\beta\uparrow,a\alpha\uparrow} - E_{b\beta\downarrow,a\alpha\downarrow}$, using on-site and Slater-Koster parameters fitted independently for the spin-up and spin-down states in the absence of spin-orbit coupling.\cite{12n_shi:2004_a}
Using the averages of spin-up/down parameters (non-magnetic structure with $\varepsilon_{a\alpha}$ and $E_{b\beta,a\alpha}$) and their differences (Stoner parameters) allows us to rotate the magnetization direction according to Eq.~(\ref{Hfm}) when the spin-orbit coupling is considered.

In order to account for the MAE we add the spin-orbit coupling term $H_{SO}$ in its atomic $\bk$-independent form to the on-site terms of our Hamiltonian:
\begin{equation} \label{Hso}
\langle b\beta\zeta\bk|H_{SO}|a\alpha\xi\bk\rangle = \delta_{b,a}\lambda_{a\beta,a\alpha} {\bm L}_{\beta\alpha}\cdot{\bm S}_{\zeta\xi}
\end{equation}
where $\lambda_{a\beta,a\alpha}$ is the spin-orbit parameter for orbitals $\alpha$, $\beta$ ($p$ does not couple with $d$) and site $a$, ${\bm L}$ is the orbital momentum operator, and ${\bm S}$ is the spin operator. The matrix elements of $H_{SO}$ in the basis of cubic harmonics are given in Eqs.~(\ref{app_Hso_p}) and~(\ref{app_Hso_d}) in the Appendix. 

Finally, we do not introduce explicitly any Hamiltonian term controlling the charge transfer between sites occupied by different atoms. Prior to our calculation we shift all on-site energies of Co with respect to Pt so that the Fermi energies calculated for pure Co and pure Pt are equal. We check that the local charges on Pt sites and on Co sites in CoPt $L1_0$ structure are in agreement with the LAPW results within the error-bar caused by charge located outside of atomic spheres used by the \textit{ab initio} method.

\section{Parametrization}
\label{se_parameters}
As mentioned above, our model relies on input parameters that are obtained by fitting band structures and total energies to \textit{ab initio} results. Extensive parameter sets~\cite{12n_papaconstantopoulos:2003_a} are available for bulk single-element metals. They assume non-orthogonal basis set, interaction to higher order neighbours, and reproduce \textit{ab initio} data with great accuracy. Our aim is to study TB models of more complex systems such as ordered binary alloys (this work) and heterostructures (future work) suitable for exploring the physics of  relativistic equilibrium and potentially also transport phenomena in these systems. 
Therefore we prefer smaller, more transferable sets of parameters assuming interactions only up to third nearest neighbours. 

We use a parametrisation by Shi and Papaconstantopoulos~\cite{12n_shi:2004_a} which further develops the Harrison approach.~\cite{12n_harrison:1980_a} Harrison expressed the two-centre Slater-Koster parameters $V_{\alpha\beta\gamma}(\rho)$ as functions of the inter-atomic distance $\rho=|{\bm \rho}|$, an effective radius of the $d$~orbital $r_d$ which is characteristic to each transition metal, and constants $\eta_{\alpha\beta\gamma}$ which are universal for all elements and lattice structures:
\begin{eqnarray}
\label{Harr_V}
  V_{\alpha\beta\gamma}(\rho) & = & \eta_{\alpha\beta\gamma}\frac{\hbar^2}{m\rho^2}, \nonumber \\
  V_{\alpha d\gamma}(\rho) & = & \eta_{\alpha d\gamma}\frac{\hbar^2r_d^{3/2}}{m\rho^{7/2}}, \nonumber \\
  V_{dd\gamma}(\rho) & = & \eta_{dd\gamma}\frac{\hbar^2r_d^3}{m\rho^5}, \;
\end{eqnarray}
where $\alpha$ and $\beta$ run through the orbitals $s$ and $p$. Values of $\eta_{\alpha\beta\gamma}$ are listed in Ref.~[\onlinecite{12n_shi:2004_a}]. In case of transition metals, Harrison used only the $s$ and $d$~orbitals so there were only two parameters specific to each element, the $d$-band width given by $r_d$ and the on-site energy of the $d$~orbitals with respect to the $s$~orbitals. 

Shi and Papaconstantopoulos significantly improved the ability of the Harrison parametrisation to produce accurate numerical results for the band structure 
while keeping the form and universality of the Slater-Koster parameters given in Eq.~(\ref{Harr_V}).
This is accomplished by: 1)~Replacing the atomic energies by on-site energies fitted to Augmented Plain Wave (APW) calculations; 2)~Including the $p$~orbitals into the basis set; 3)~Modifying of the $sp$~Slater-Koster parameters by introducing a dimensionless parameter $\gamma_s$ as follows:
\begin{equation}
\label{gamma_s}
  V_{\alpha\beta\gamma}(\rho) = \eta_{\alpha\beta\gamma}\frac{\gamma_s \hbar^2}{m\rho^2};
\end{equation}
4)~Obtaining new prefactors $\eta_{\alpha\beta\gamma}$ and radii $r_d$ by simultaneously fitting the APW energy bands of 12 transition metals at the equilibrium lattice constants of the particular element. The new parameters reproduced APW energy bands and density of states (DOS) remarkably well, not only for the 12 elements originally fitted, but also for the rest of the transition metals, the alkaline earth and the noble metals as shown in Ref.~[\onlinecite{12n_shi:2004_a}]. This parametrisation assumes an orthogonal basis set and interaction to second (fcc) or third (bcc) nearest neighbours. The on-site and Slater-Koster parameters are exchange-split in case of the ferromagnetic metals.

In this work we build on the results of Shi and Papaconstantopoulos and test the transferability of their TB parameters to an ordered binary alloy. We use the parameters for single-element bulk metals\cite{12n_shi:2004_a} with the following modifications: 1)~The interatomic Slater-Koster parameters between Co and Pt atoms are set to a geometric average of the elemental values: $V^{Co,Pt}_{\alpha\beta\gamma}= \sqrt{V^{Co}_{\alpha\beta\gamma}V^{Pt}_{\alpha\beta\gamma}}$ following the work of Ballhausen and Gray.\cite{ballhausen:1964_a} Using the geometric average is also in line with the LMTO\cite{Skriver:1984_a} method; 2)~The exchange-splittings of the on-site and Slater-Koster parameters enter our model through the on-site ($I_{a\alpha}$) and hopping ($I_{a\alpha,b\beta}$) Stoner parameters, respectively. We increase $I_{Co,d}$ and introduce non-zero $I_{Pt,d}$ to reproduce the LAPW spin magnetic moment and DOS of CoPt more accurately; 3)~We vary the on-site energy of unoccupied $p$~orbitals ($\varepsilon_{Pt,p}$), which extend the original Harrison's set of parameters and are specific to a particular single-element crystal, in order to further improve the agreement with spin magnetic moment and DOS obtained by LAPW and explore the dependence of MAE in the CoPt compound on this parameter.

Finally, we add atomic spin-orbit coupling parameters obtained by numerical Hartree-Fock calculations based on the Dirac-Coulomb Hamiltonian\cite{12n_visscher:1997_a} to the Pt sites. The $5d$~orbital has an atomic value $\lambda_{Pt,d}=0.0445$~Ry and the $3d$~orbital of Co has $\lambda_{Co,d}=0.0063$~Ry. We neglect the spin-orbit coupling in $6p$-orbitals of Pt and $4p$-orbitals of Co to keep the number of input parameters low. 

\section{Comparison of Tight Binding and \textit{Ab Initio} results}
\label{se_comparison}
We check the validity of our extensions of the Harrison TB model and the transferability of Shi and Papaconstantopoulos' parametrization to CoPt ordered alloy by comparing the band structure, spin magnetic moment, DOS, and MAE calculated using our TB code and a well established \textit{ab initio} code, the full-potential relativistic LAPW package Wien2K.\cite{12n_blaha:1990,12n_Schwartz:2003_a}

In CoPt $L1_0$ structure, fcc lattice sites are occupied by alternating layers of Co and Pt atoms and the lattice constant perpendicular to layers $c$ is smaller than the in-plane lattice constant $a$. Throughout this work we use the experimental lattice constants: $a=7.19$~a.u, $c=7.01$~a.u.\cite{12n_shick:2003_a}

The Slater-Koster parameters for single-element fcc crystals\cite{12n_shi:2004_a} were fitted assuming interaction to second nearest neighbours. However, we intend to account for MAE caused by hybridisation of magnetic $3d$~orbitals on Co and spin-orbit coupled $5d$~orbitals on Pt. The coupling between the Co and Pt sites in the $L1_0$ lattice in our two-center approximation is formed by 8 (16) second (third) nearest neighbour hopping integrals. Therefore, to enhance the ability of our model to capture the Co-Pt hybridisation, we add the third nearest neighbours to the sum in Eq.~(\ref{ham_m_band}).

\begin{figure}
\centering
\begin{tabular}{c}
\includegraphics[width=0.99\columnwidth]{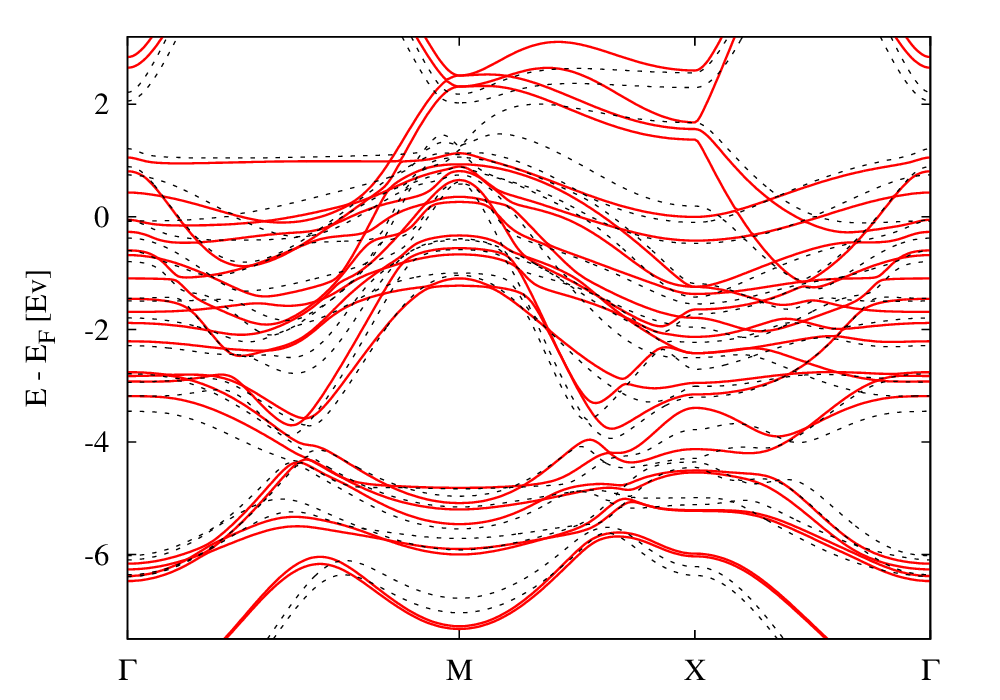} \\
(a) Second nearest neighbours \vspace*{0.2cm} \\
\includegraphics[width=0.99\columnwidth]{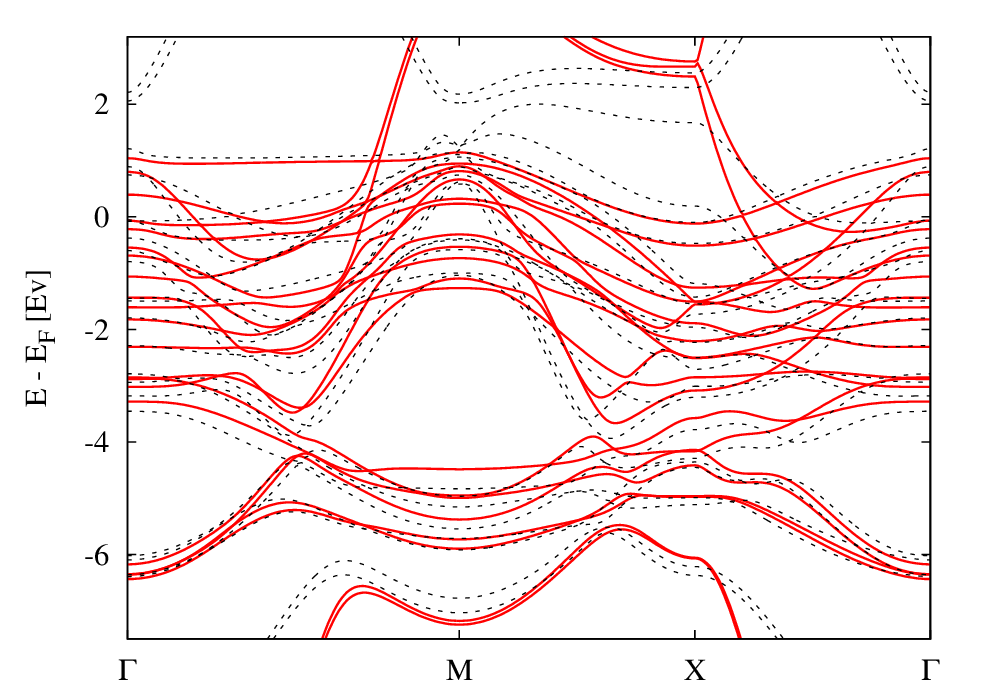} \\
(b) Third nearest neighbours \\
\end{tabular}
\caption{(Color online) Comparison of TB (continuous red lines) and LAPW-LSDA (dashed black lines) band structures of CoPt $L_10$ structure with magnetization along the $[001]$ axis. Second (a) and third (b) nearest neighbours together with ``atomic parameters'' are assumed in TB whereas LAPW bands are the same in both plots.}
\label{f_bands}
\end{figure}

Fig.~\ref{f_bands} shows band structures calculated using our TB model assuming second (a) and third (b) nearest neighbours compared to LAPW results with local spin density approximation (LSDA) of the exchange-correlation potential. At this stage we do not modify the TB parameters derived for single-element metals except taking geometric average of Slater-Koster parameters between Co and Pt. We denote this default set of TB parameters as the ``atomic parameters''. 

There is a good agreement of the bands especially in the vicinity of the Fermi energy. TB bands assuming second nearest neighbours match LAPW bands also in the region of $s$ and $p$~states. However, the bottom of the $s$-band is very far from the Fermi energy and the $p$-band has only an auxiliary role in our model so we can conclude that summation to both second and third neighbours has potential for successful simulation of the MAE in CoPt $L1_0$ structure.

Now we turn our attention to the spin magnetic moment, spin-resolved DOS and MAE calculated using second nearest neighbours together with the ``atomic parameters'' and compare these TB results to LAPW with LSDA and spin-polarised generalized gradient approximation (GGA). It is known that GGA and LDA are suitable for $3d$ and $5d$ transition metals, respectively, so it is not clear which approximation is more appropriate for the CoPt compound. Therefore, we try to view the difference between TB and LSDA results in the context of the difference between calculations using LSDA and GGA approximations. 

\begin{figure}
\includegraphics[width=0.97\columnwidth]{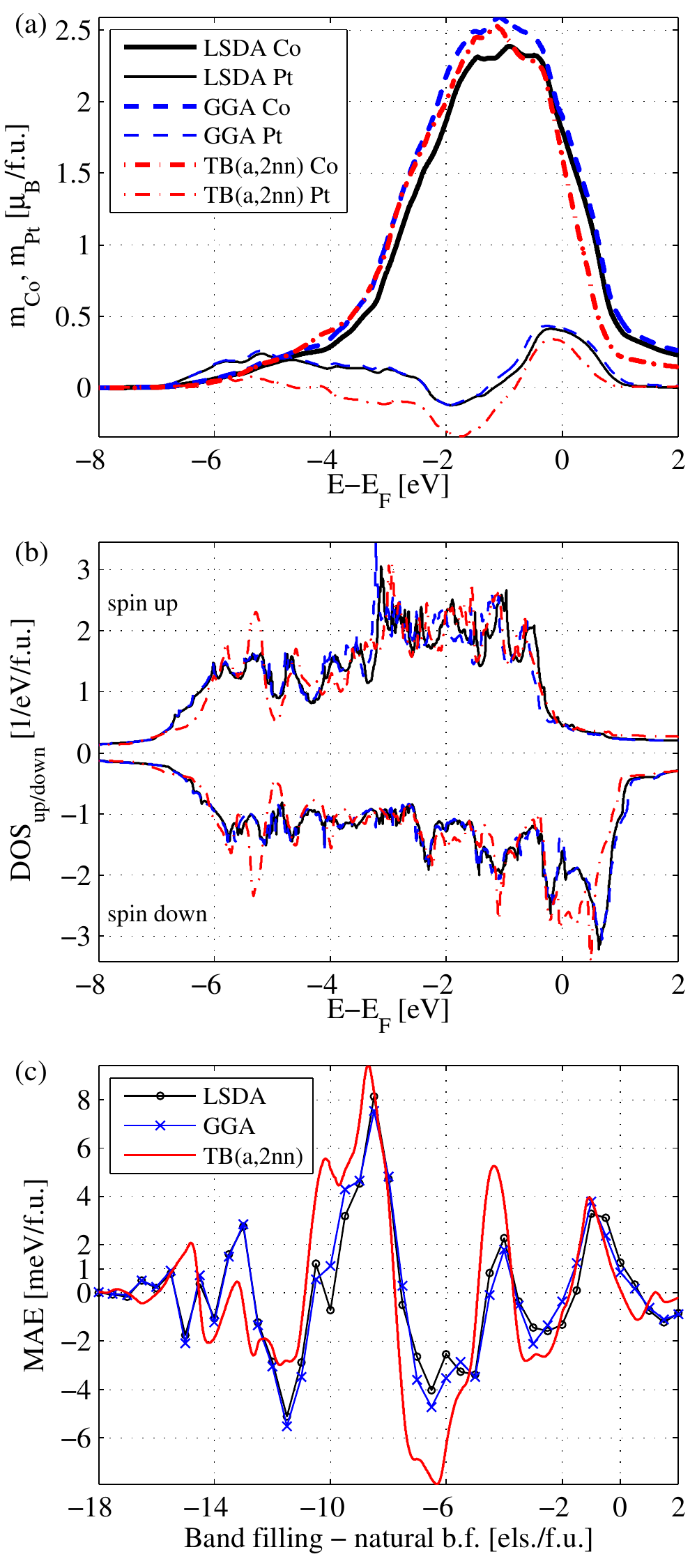}
\caption{(Color online) Comparison of TB results assuming ``atomic parameters'' and second nearest neighbours (a,2nn) to LAPW using LSDA and GGA approximations for quantities: (a)~Spin magnetic moment on Co (thick lines) and Pt (thin lines) sites per formula unit with magnetization along the $[001]$ axis (spin in the interstitial region of LAPW is not captured); (b)~DOS  per formula unit projected on spin-up/down states with magnetization along the $[001]$ axis, the legend of plot (a) applies including the line style used for Pt and color coding; (c)~MAE~$=E_{tot}(M_{110})-E_{tot}(M_{001})$ per formula unit for a range of band fillings where the natural band filling is 19 valence electrons.}
\label{f_cmp_param_wa}
\end{figure}

Fig.~\ref{f_cmp_param_wa}(a) shows the spin magnetic moment on Co and Pt sites per formula unit (f.u.) for energies ranging through the whole valence band for magnetization along the $[001]$ axis. (Our $[001]$ axis is set perpendicular to the alternating Co and Pt atomic planes and the nearest in-plane neighbour lies on the $[110]$ axis.) The Wien2K code places all atoms of the unit cell in non-overlapping spheres leaving some charge in the interstitial region. We set the radii of the atomic spheres to $r_{Co}=2.2$~a.u. and $r_{Pt}=2.4$~a.u. for all LAPW calculations causing our spin magnetic moment in the interstitial region to be less than $5\%$ of the total spin magnetic moment. GGA and LSDA give very similar values in case of Pt and deviate slightly in case of Co in the middle of the $d$-band. TB spin magnetic moment on Pt is lower than the LAPW prediction in the whole energy range, whereas TB values for Co match the GGA very well except around the Fermi level where Co is predicted to be less spin-polarized by TB than by LSDA and GGA. The total spin magnetic moment is $2.27$~$\mu_B/$f.u. using GGA and $2.21$~$\mu_B/$f.u. using LSDA which is in good agreement with the experimental value $2.4$~$\mu_B/$f.u. TB prediction of the total spin magnetic moment is $1.92$~$\mu_B/$f.u. Very similar results are obtained for magnetization along the $[110]$ axis so we do not plot them.

Fig.~\ref{f_cmp_param_wa}(b) shows DOS projected on spin-up/down states again for magnetization along the $[001]$ axis and summation to second nearest neighbours. LSDA and GGA are in excellent agreement in case of spin-down but differ slightly in case of spin-up. The DOS calculated by TB has unexpected peaks close to the bottom of the $d$-band. A separate calculation of the Pt-component of DOS and the fact that these peaks for spin-up/down are not mutually shifted in energy suggests that they correspond to Pt-states. We hypothesize at this stage that an increased coupling between Co and Pt sites could remove these unrealistic peaks. In general, the spin-up part of the valence band calculated by TB also seems to be less shifted in energy with respect to the spin-down part as compared to the LAPW reference. This feature of DOS corresponds to the observation of lower spin magnetic moment predicted by TB both for Co and Pt at the Fermi level in Fig.~\ref{f_cmp_param_wa}(a) and motivates us to increase the Stoner parameters both for Co and Pt atoms and add the third nearest neighbours to enhance the coupling between Co and Pt in the next section.

Fig.~\ref{f_cmp_param_wa}(c) compares MAE calculated again using TB with ``atomic parameters'' and second nearest neighbours and LAPW with LSDA and GGA approximations. Note that in the TB case MAE amounts to the difference of total energies for two magnetization directions: MAE~$=E_{tot}(M_{110})-E_{tot}(M_{001})$, whereas in case of LAPW we use the force theorem
following the work of Shick.\cite{12n_shick:2003_a} (We use about 50000 and 200000 $k$-points in LAPW and TB integration, respectively.) MAE is a more subtle quantity than the spin magnetic moment or DOS so we calculate it for a wide range of band filling~(b.f.) using the rigid band approximation and compare the trends of MAE(b.f.) rather than individual values. 

We observe a very good agreement of the LSDA and GGA data and a broad agreement of the TB and \textit{ab initio} curves. We consider this a remarkable success of the TB model which employs parameters fitted for pure single-element metals. The amplitude of MAE oscillations decreases towards the edges of the valence band so the relative error of the TB result at the Fermi energy becomes quite large. Both LAPW values MAE$_{LSDA}= 1.26$~meV/f.u. and MAE$_{GGA}= 0.85$~meV/f.u. are in good agreement with the measured value MAE$_{exp.}= 1.0$~meV/f.u.,\cite{yermakov:1990_a} whereas our "atomic parameters" TB prediction based on straight forward transfer of Shi and Papaconstantopoulos' parameters is an order of magnitude lower: MAE$_{TB}=0.13$~meV/f.u. 


\section{Parameter Refinement and MAE}
\label{se_refinement}

The good agreement of key features in the spin magnetic moment, DOS, and MAE calculated by LAPW and TB motivates us to improve our TB model. The shortcomings of the initial simulations described in the previous section provide useful guidance how to proceed. As concluded in the discussion of Fig.~\ref{f_cmp_param_wa}(b), we add the third nearest neighbours to enhance the coupling between Co and Pt sites and we increase the on-site Stoner parameters both for Co and Pt atoms to match the larger spin polarization obtained by LAPW. 

\begin{figure}
\includegraphics[width=0.97\columnwidth]{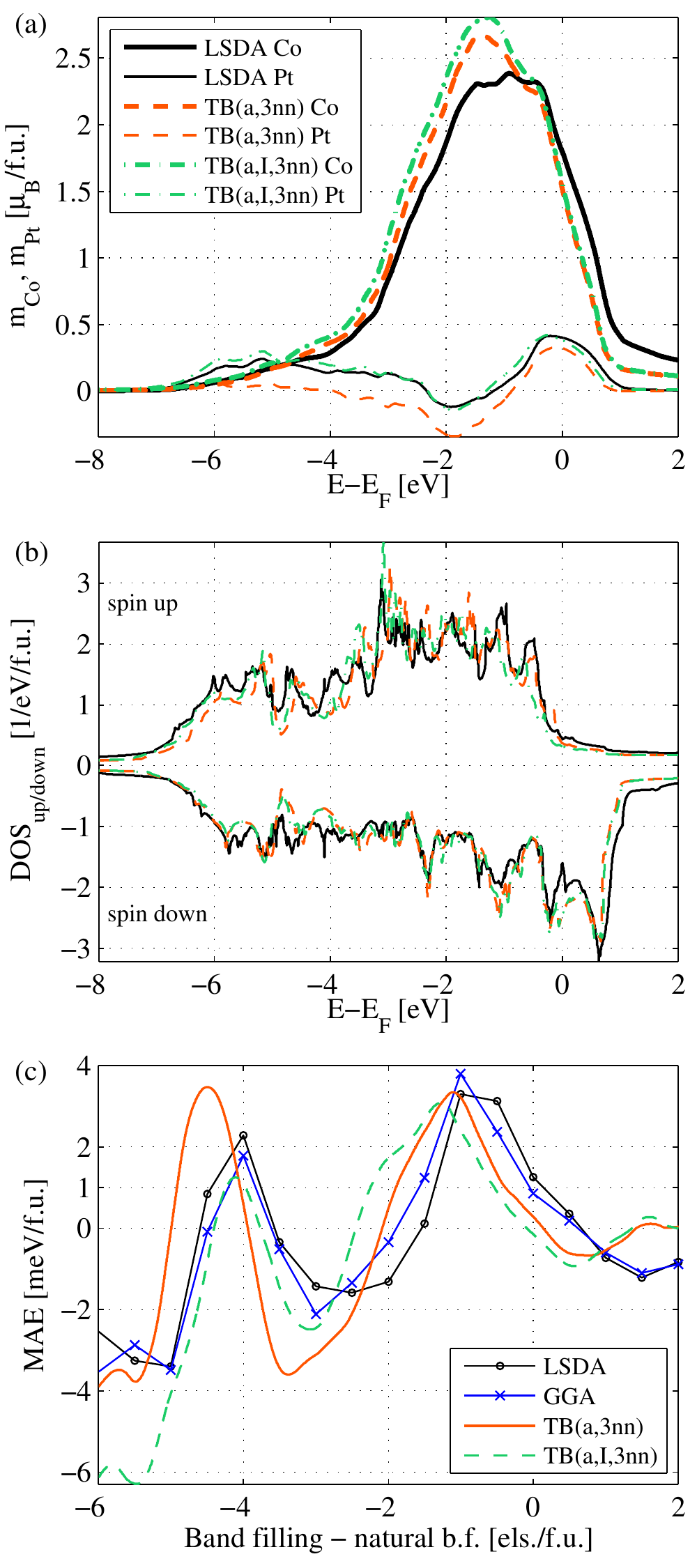}
\caption{(Color online) Comparison of TB results assuming  third nearest neighbours and ``atomic parameters'' (a,3nn) or ``atomic parameters'' with enhanced Stoner parameter (a,I,3nn) to LAPW using LSDA approximation for quantities: (a)~Spin magnetic moment on Co (thick lines) and Pt (thin lines) sites with magnetization along the $[001]$ axis; (b)~DOS projected on spin-up/down states with magnetization along the $[001]$ axis, the legend of plot (a) applies including the line style used for Pt and color coding; (c)~MAE~$=E_{tot}(M_{110})-E_{tot}(M_{001})$ for a range of band fillings where the natural band filling is 19 valence electrons.}
\label{f_cmp_param_ae}
\end{figure}

The results are summarised in Fig.~\ref{f_cmp_param_ae} which presents the same quantities as Fig.~\ref{f_cmp_param_wa} computed in three different ways: TB with ``atomic parameters'' and third nearest neighbours (a,3nn); TB with ``atomic parameters'' enhanced by $I_{Co,d}=1.08I_{Co,d}^{atomic}=0.134$~Ry, $I_{Pt,d}=0.15I_{Co,d}$, and third nearest neighbours (a,I,3nn); LAPW with LSDA (same as in Fig.~\ref{f_cmp_param_wa}, GGA is not included to maintain legibility of the plots).

Fig.~\ref{f_cmp_param_ae}(a) can be contrasted with its counterpart, Fig.~\ref{f_cmp_param_wa}(a). The spin magnetic moment assuming the third nearest neighbours is again underestimated by TB using the default ``atomic parameters''. However, we managed to tune the Stoner parameters $I_{Pt,d}$ and $I_{Co,d}$ to achieve quantitative agreement of the spin magnetic moment on Pt with the LSDA values throughout the whole valence band. The spin magnetic moment on Co cannot be brought to a full quantitative agreement with LSDA by tuning only parameters $I_{Pt,d}$ and $I_{Co,d}$ and remains slightly lower than the LSDA or GGA reference at the Fermi energy. We address this deficit in the final refinement of our TB model.

DOS in Fig.~\ref{f_cmp_param_ae}(b) shows significant improvement over Fig.~\ref{f_cmp_param_wa}(b) so the summation to third nearest neighbours seems to be more suitable for modelling the coupling between Co and Pt. Moreover, the enhancement of the Stoner parameter increased the mutual shift of spin-up/down DOS to show closer match with LSDA, as expected. 

In Fig.~\ref{f_cmp_param_ae}(c) we focus on band fillings closer to the natural band filling (b.f.~=~19 valence electrons). Including the third nearest neighbours does not cause any significant change in the overall MAE dependence on the band filling. Slightly more pronounced change corresponds to increasing the Stoner parameters but the main features of MAE are still in agreement with LSDA and GGA data. Note that the MAE at the natural band filling becomes negative due to the enhanced exchange interaction.

We can conclude that our first TB parameter refinement attempt presented in Fig.~\ref{f_cmp_param_ae} leads to a better agreement with LAPW in a broad range of valence band energies, however, increasing $I_{Pt,d}$ and $I_{Co,d}$ and adding third nearest neighbours does not reproduce accurately the net spin magnetic moment predicted by LAPW at the Fermi energy and the MAE at the natural band filling.

As we mentioned in Secs.~\ref{se_parameters} and~\ref{se_comparison} the unoccupied $p$-states were added to the Harrison TB model by Shi and Papaconstantopoulos only to produce more realistic warping of the $d$-band in single-element crystals close to Fermi energy. Therefore, the position of the unoccupied $p$-states in the CoPt compound is the next natural subject to scrutiny. 

We note that Pt offers more room for variation of the $p$-state on-site energy as its $s$-state is much lower in energy than the $s$-state of Co: $(\varepsilon_{Pt,p}-\varepsilon_{Pt,d})/ (\varepsilon_{Pt,s}-\varepsilon_{Pt,d})=2.2$ whereas $(\varepsilon_{Co,p}-\varepsilon_{Co,d})/ (\varepsilon_{Co,s}-\varepsilon_{Co,d})=1.02$. Therefore, we can bring $\varepsilon_{Pt,p}$ closer to $\varepsilon_{Pt,s}$ (and to $\varepsilon_{Pt,d}$ further below) without changing the order of the Pt on-site energies. Such shift should increase the hybridisation of the $p$-states on Pt with the exchange-split $d$-states increasing the deficient net spin magnetic moment.

\begin{figure}
\includegraphics[width=0.97\columnwidth]{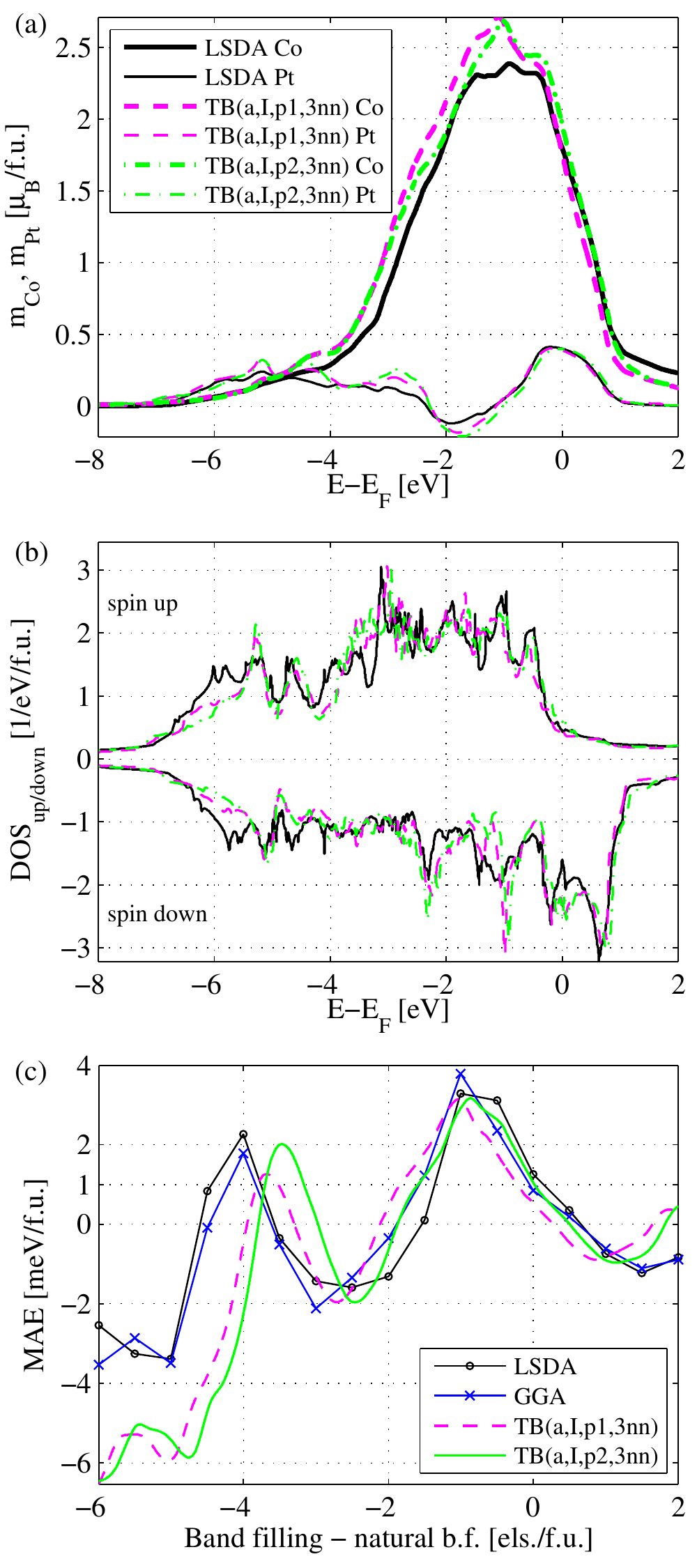}
\caption{(Color online) Comparison of TB results assuming  third nearest neighbours, ``atomic parameters'' with enhanced Stoner parameter, and on-site energies $\varepsilon_{Pt,p}-\varepsilon_{Pt,d}=0.69$~Ry (a,I,p1,3nn) or $\varepsilon_{Pt,p}-\varepsilon_{Pt,d}=0.54$~Ry (a,I,p2,3nn) to LSDA approximation for quantities: (a)~Spin magnetic moment on Co (thick lines) and Pt (thin lines) sites with magnetization along the [001] axis; (b)~DOS projected on spin-up/down states with magnetization along the [001] axis, the legend of plot (a) applies including the line style used for Pt and color coding; (c)~MAE~$=E_{tot}(M_{110})-E_{tot}(M_{001})$ for a range of band fillings where the natural band filling is 19 valence electrons.}
\label{f_cmp_param_pp}
\end{figure}

Fig.~\ref{f_cmp_param_pp} shows an overview of TB and LAPW results analogous to Fig.~\ref{f_cmp_param_ae}. The new TB data are calculated using ``atomic parameters'' with enhanced Stoner parameter, third nearest neighbours, and two examples of the shifted Pt on-site energies: $(\varepsilon_{Pt,p}-\varepsilon_{Pt,d})/ (\varepsilon_{Pt,s}-\varepsilon_{Pt,d})=1.61$ 
labelled as (a,I,p1,3nn) and 
$(\varepsilon_{Pt,p}-\varepsilon_{Pt,d})/ (\varepsilon_{Pt,s}-\varepsilon_{Pt,d})=1.26$ 
labelled as (a,I,p2,3nn). The LSDA data are the same as in Figs.~\ref{f_cmp_param_wa} and~\ref{f_cmp_param_ae}.

As expected, the greater proximity of $p$~and $d$-states increases the spin polarization of Pt deeper in the valence band, however, the effects compensate at the Fermi energy giving excellent agreement with LSDA data as shown in Fig.~\ref{f_cmp_param_pp}(a). At the same time, the spin magnetic moment of Co at Fermi energy scales with the shift of on-site energy $\varepsilon_{Pt,p}$ due to hybridisation with Pt and finally reaches the LAPW values when the shift mentioned above is in the range $\varepsilon_{Pt,p}-\varepsilon_{Pt,d}\approx0.54-0.69$~Ry. 

The agreement of DOS calculated by LAPW and TB with enhanced Stoner parameters shown in Fig.~\ref{f_cmp_param_ae}(b) is satisfactory. We include Fig.~\ref{f_cmp_param_pp}(b) to demonstrate that shifting $\varepsilon_{Pt,p}$ causes only minor deviation from DOS obtained by LSDA at the very bottom of the valence band, whereas it further improves the agreement closer to the Fermi energy.

Fig.~\ref{f_cmp_param_pp}(c) shows the main result of our work. The general trend of the MAE dependence on the band filling is very robust against small variations of the TB input parameters and is in broad agreement with MAE calculated by LAPW. On the other hand, the value of MAE for the natural band filling turns out to be very sensitive to the input parameters. Remarkably, the MAE for the natural band filling is in quantitative agreement with the LSDA, GGA, and experimental values MAE~$\approx 1$~meV/f.u. when the on-site energy of the $p$-states on Pt is in the range determined by comparing DOS and spin magnetic moment to LAPW: $\varepsilon_{Pt,p}-\varepsilon_{Pt,d}\approx0.54-0.69$~Ry.

We have also explored the sensitivity of the above quantities to the variation of the Slater-Koster parameters. However, replacing the geometric average by an arithmetic average to obtain $V^{Co,Pt}_{\alpha\beta\gamma}$ or increasing the relative magnitude of the Slater-Koster parameters to enhance the Co-Pt hybridisation does not change the spin magnetic moment or DOS in a transparent way that would improve our physical understanding of the electronic structure or the overall agreement with the LAPW results.

\section{Summary}
\label{se_summary}
We have carried out systematic modeling of electronic structure and relativistic magnetic characteristics of bulk CoPt $L1_0$ structure using TB and \textit{ab initio} methods. An emprirical multiorbital TB model following the Harrison approach with a parametrisation devised by Shi and Papaconstantopoulos was applied. We extended the model by adding an atomic spin-orbit coupling term to the on-site Hamiltonian blocks in order to account for the magnetocrystalline anisotropy. We have focused on the MAE as a function of the band filling so that we could compare general trends of TB and LAPW rather than singe values for the natural band filling which are available in literature.

We started our calculations by checking the validity of the model assuming the ``atomic parameters'' (parameters optimized for single-element fcc crystals with geometric averaging of Slater-Koster parameters between Co and Pt) by comparing the TB band structure, spin magnetic moment, DOS, and MAE to corresponding LAPW results obtained using the Wien2K program package. The broad agreement observed throughout the valence band with slightly deficient TB net spin magnetic moment and smaller mutual shift of the projected DOS stimulated further development of the parametrisation focusing on the Stoner parameters and on-site energies of the virtual $p$-states.

We continued by adding the interaction to third nearest neighbours
to enhance the hybridisation between magnetic and spin-orbit coupled sites and varied three TB parameters ($I_{Pt,d}$, $I_{Co,d}$, and $\varepsilon_{Pt,p}$) which extend the original Harrison parametrisation and their values were likely to require corrections after the transfer to an ordered binary alloy. We compared the new set of results to \textit{ab initio} predictions again. In other works, this process is typically replaced by simultaneous fitting of all TB parameters to APW band structures and total energies, however, in our work we seek better physical insight into the spin-orbit coupling phenomena and transferability between different structures rather than precise agreement of TB and \textit{ab initio} results. 

We found that the net spin magnetic moment increases with small enhancement of the Stoner parameters and with shifting of the Pt $p$-state on-site energy towards the $d$-states as expected. Remarkably, the MAE obtained for a narrow range of these parameters, where the net spin magnetic moment and DOS reached the best agreement with the \textit{ab initio} predictions, is in quantitative agreement with \textit{ab initio} results.  
Such success motivates future investigations of the transferability of the model in other compounds or multilayers. At the same time our TB model is well suited for incorporation of the equilibrium Green's function framework to calculate relativistic magnetotransport phenomena in structures containing the magnetic compounds or multilayers.

\acknowledgments
We acknowledge fruitful discussions with  A.~W.~Rushforth, K.~V\'{y}born\'{y}, Jairo Sinova, and V.~Amin and support from the EPSRC Grant No. EP/H029257/1, from the EU ERC Advanced Grant No. 268066, from the Ministry of Education of the Czech Republic Grant No. LM2011026, and from the Academy of Sciences of the Czech Republic Preamium Academiae.

\appendix

\section{Spin-Orbit Hamiltonian}
\label{ase_SO}
The spin-orbit Hamiltonian term has to be written in the basis of cubic harmonics for which the Slater-Koster parameters are derived in literature.\cite{12n_slater:1954_a,12n_shi:2004_a} The cubic harmonics can be written in terms of the angular momentum eigenstates for the $s$, $p$, and $d$~orbitals:
\begin{eqnarray}\label{app_charm}
s_0 &=& Y_0^0, \nonumber \\
p_x &=&\frac{1}{\sqrt{2}}\big(Y_1^{-1} - Y_1^1 \big), \nonumber \\ 
p_y &=&\frac{i}{\sqrt{2}}\big(Y_1^{-1} + Y_1^1 \big), \nonumber \\ 
p_z &=& Y_1^0, \nonumber \\ 
d_{xy} &      =&\frac{i}{\sqrt{2}}\big( Y_2^{-2}  -  Y_2^2 \big), \nonumber \\ 
d_{yz} &      =&\frac{i}{\sqrt{2}}\big( Y_2^{-1}  +  Y_2^1 \big), \nonumber \\ 
d_{xz} &      =&\frac{1}{\sqrt{2}}\big( Y_2^{-1}  -  Y_2^1 \big), \nonumber \\
d_{x^2-y^2} & =&\frac{1}{\sqrt{2}}\big( Y_2^{-2}  +  Y_2^2 \big), \nonumber \\
d_{3z^2-r^2} &=& Y_2^0,  \
\end{eqnarray}
where $\langle {\bm \hat{n}}|l,m\rangle = Y_l^m(\theta,\phi)$, $L_z|l,m\rangle = m|l,m\rangle$, and ${\bm L}^2|l,m\rangle = l(l+1)|l,m\rangle$. In atomic units the angular and spin moments are dimensionless and the coefficient $\lambda_{a\alpha}$ in Eq.~(\ref{Hso}) is measured in Rydbergs. 
With cubic harmonics ordered as in Eq.~(\ref{app_charm}) we obtain the following Hamiltonian contribution for the $p$~orbitals:
\begin{equation}\label{app_Hso_p}
H_{SO,p} = \frac{\lambda_p}{2} \left[ \begin{array}{ccc|ccc}
 0 & -i &  0   &   0 & 0 &  1 \\
 i &  0 &  0   &   0 & 0 & -i \\
 0 &  0 &  0   &  -1 & i &  0 \\ \hline
 0 &  0 & -1   &   0 & i &  0 \\
 0 &  0 & -i   &  -i & 0 &  0 \\
 1 &  i &  0   &   0 & 0 &  0 \\
\end{array} \right],
\end{equation}
and for the $d$~orbitals:
\begin{widetext}
\begin{equation}\label{app_Hso_d}
H_{SO,d} = \frac{\lambda_d}{2} \left[ \begin{array}{ccccc|ccccc}
  0 &  0 & 0 & 2i & 0   &   0 & 1 & -i &  0 &  0               \\
  0 &  0 & i &  0 & 0   &  -1 & 0 &  0 & -i & -i\sqrt{3}       \\
  0 & -i & 0 &  0 & 0   &   i & 0 &  0 & -1 &  \sqrt{3}        \\
-2i &  0 & 0 &  0 & 0   &   0 & i &  1 &  0 &  0               \\
  0 &  0 & 0 &  0 & 0   &   0 & i\sqrt{3} &  -\sqrt{3} & 0 & 0 \\ \hline
 0 & -1 & -i &  0 & 0           &    0 & 0 &  0 & -2i & 0     \\
 1 &  0 &  0 & -i & -i\sqrt{3}  &    0 & 0 & -i &   0 & 0     \\
 i &  0 &  0 &  1 & -\sqrt{3}   &    0 & i &  0 &   0 & 0     \\
 0 &  i & -1 &  0 & 0           &   2i & 0 &  0 &   0 & 0     \\
 0 &  i\sqrt{3} & \sqrt{3} & 0  &    0 & 0 &  0 &   0 & 0 & 0 \\ 
\end{array} \right] \nonumber \;
\end{equation}
\end{widetext}
where the first and second diagonal blocks correspond to spin up and down eigenstates: $S_z|l_s,m_s \rangle=\pm1/2|l_s,m_s \rangle$, respectively. The total $H_{SO}$ matrix is added to the on-site Hamiltonian terms, whereas the inter-site matrix elements remain unchanged. The size of $\lambda_{a\alpha}$ depends on the type of atom. We neglect $\lambda_{Co,p}$ and $\lambda_{Pt,p}$ as they are at least an order of magnitude smaller than $\lambda_{Co,d}$ and $\lambda_{Pt,d}$.

\section{Spin-Orbit Hamiltonian}
\label{ase_SK}
As mentioned in Sec.~\ref{se_model} the hopping integrals $E_{b\beta,a\alpha}({\bm \rho}_n)$ can be written in terms of Slater-Koster\cite{12n_slater:1954_a} parameters $V^{b,a}_{\beta\alpha\upsilon}$ (neglecting the site indices) for the $s$ and $p$~orbitals:
\begin{eqnarray}\label{app_E_Vsp}
  E_{s,s} & = & V_{ss\sigma}, \nonumber \\
  E_{s,x} & = & lV_{sp\sigma}, \nonumber \\
  E_{x,x} & = & l^2V_{pp\sigma}+(1-l^2)V_{pp\pi}, \nonumber \\
  E_{x,y} & = & lmV_{pp\sigma}-lmV_{pp\pi}, \nonumber \\
  E_{x,z} & = & lnV_{pp\sigma}-lnV_{pp\pi}, \;
\end{eqnarray}
and for the $s$, $p$, and $d$~orbitals:
\begin{widetext}
\begin{eqnarray}\label{app_E_Vd}
  E_{s,xy} & = & \sqrt{3}lmV_{sd\sigma}, \nonumber \\
  E_{s,x^2-y^2} & = & \frac{\sqrt{3}}{2}(l^2-m^2)V_{sd\sigma}, \nonumber \\
  E_{s,3z^2-r^2} & = & [n^2-\frac{1}{2}(l^2+m^2)]V_{sd\sigma}, \nonumber \\
  E_{x,xy} & = & \sqrt{3}l^2mV_{pd\sigma} + m(1 - 2l^2)V_{pd\pi}, \nonumber \\
  E_{x,yz} & = & \sqrt{3}lmnV_{pd\sigma} - 2lmnV_{pd\pi}, \nonumber \\
  E_{x,zx} & = & \sqrt{3}l^2nV_{pd\sigma} + n(1 - 2l^2)V_{pd\pi}, \nonumber \\
  E_{x,x^2-y^2} & = & \frac{\sqrt{3}}{2}l(l^2-m^2)V_{pd\sigma} + l(1-l^2+m^2)V_{pd\pi}, \nonumber \\
  E_{y,x^2-y^2} & = & \frac{\sqrt{3}}{2}m(l^2-m^2)V_{pd\sigma} - m(1+l^2-m^2)V_{pd\pi}, \nonumber \\
  E_{z,x^2-y^2} & = & \frac{\sqrt{3}}{2}n(l^2-m^2)V_{pd\sigma} - n(l^2-m^2)V_{pd\pi}, \nonumber \\
  E_{x,3z^2-r^2} & = & l[n^2 - \frac{1}{2}(l^2 + m^2)]V_{pd\sigma} - \sqrt{3} l n^2 V_{pd\pi}, \nonumber \\
  E_{y,3z^2-r^2} & = & m[n^2 - \frac{1}{2}(l^2 + m^2)]V_{pd\sigma} - \sqrt{3} m n^2 V_{pd\pi}, \nonumber \\
  E_{z,3z^2-r^2} & = & n[n^2 - \frac{1}{2}(l^2 + m^2)]V_{pd\sigma} + \sqrt{3} n (l^2 + m^2) V_{pd\pi}, \nonumber \\
  E_{xy,xy} & = & 3 l^2 m^2 V_{dd\sigma} + (l^2 + m^2 - 4 l^2 m^2) V_{dd\pi} + (n^2 + l^2 m^2) V_{dd\delta}, \nonumber \\
  E_{xy,yz} & = & 3 l m^2 nV_{dd\sigma} + l n (1 - 4 m^2) V_{dd\pi} + l n (m^2 - 1) V_{dd\delta}, \nonumber \\
  E_{xy,zx} & = & 3 l^2 m n V_{dd\sigma} + m n (1 - 4 l^2) V_{dd\pi} + m n (l^2 - 1) V_{dd\delta}, \nonumber \\
  E_{xy,x^2-y^2} & = & \frac{3}{2} l m (l^2 - m^2) V_{dd\sigma} + 2 l m (m^2 - l^2) V_{dd\pi} + \frac{1}{2}l m (l^2 - m^2) V_{dd\delta}, \nonumber \\
  E_{yz,x^2-y^2} & = & \frac{3}{2} m n (l^2 - m^2) V_{dd\sigma} - m n [1 + 2(l^2 - m^2)] V_{dd\pi} + m n [1 + \frac{1}{2}(l^2 - m^2)] V_{dd\delta}, \nonumber \\
  E_{zx,x^2-y^2} & = & \frac{3}{2} n l (l^2 - m^2) V_{dd\sigma} + n l [1 - 2(l^2 - m^2)] V_{dd\pi} - n l [1 - \frac{1}{2}(l^2 - m^2)] V_{dd\delta}, \nonumber \\
  E_{xy,3z^2-r^2} & = & \sqrt{3} \left[ l m (n^2 - \frac{1}{2}(l^2 + m^2)) V_{dd\sigma} - 2 l m n^2 V_{dd\pi} + \frac{1}{2} l m (1 + n^2)  V_{dd\delta} \right], \nonumber \\
  E_{yz,3z^2-r^2} & = & \sqrt{3} \left[ m n (n^2 - \frac{1}{2}(l^2 + m^2)) V_{dd\sigma} + m n (l^2 + m^2 - n^2) V_{dd\pi} - \frac{1}{2} m n (l^2 + m^2) V_{dd\delta} \right], \nonumber \\
  E_{zx,3z^2-r^2} & = & \sqrt{3} \left[ l n (n^2 - \frac{1}{2}(l^2 + m^2)) V_{dd\sigma} + l n (l^2 + m^2 - n^2) V_{dd\pi} - \frac{1}{2} l n (l^2 + m^2) V_{dd\delta} \right], \nonumber \\
  E_{x^2-y^2,x^2-y^2} & = & \frac{3}{4} (l^2 - m^2)^2 V_{dd\sigma} + [l^2 + m^2 - (l^2 - m^2)^2] V_{dd\pi} + [n^2 + \frac{1}{4}(l^2 - m^2)^2] V_{dd\delta}, \nonumber \\
  E_{x^2-y^2,3z^2-r^2} & = & \sqrt{3} \left[\frac{1}{2}(l^2 - m^2) [n^2 - \frac{1}{2}(l^2 + m^2)] V_{dd\sigma} + n^2 (m^2 - l^2) V_{dd\pi} + \frac{1}{4}(1 + n^2)(l^2 - m^2) V_{dd\delta}\right], \nonumber \\
  E_{3z^2-r^2,3z^2-r^2} & = & [n^2 - \frac{1}{2}(l^2 + m^2)]^2 V_{dd\sigma} + 3 n^2 (l^2 + m^2) V_{dd\pi} + \frac{3}{4} (l^2 + m^2)^2 V_{dd\delta}, \;
\end{eqnarray}
\end{widetext}
where $l$, $m$, and $n$ are the directional cosines of the intersite position vector ${\bm \rho}_n$ and the labels $s$, $x$, $y$, $z$, $xy$, etc. denote the cubic harmonics written in terms of spherical harmonics in Eq.~(\ref{app_charm}).

\end{document}